\PassOptionsToPackage{table,xcdraw}{xcolor}
\documentclass{article}
\usepackage{ijcai18}

\usepackage{times}
\usepackage{xcolor}
\usepackage{soul}
\usepackage[utf8]{inputenc}
\usepackage[small]{caption}

\usepackage{booktabs} 
\usepackage{tabularx}
\usepackage{upgreek}
\usepackage{amsmath}
\usepackage{bm}      
\usepackage{amsfonts}
\usepackage[ruled,vlined,linesnumbered,noresetcount]{algorithm2e}
\usepackage{multirow}
\usepackage{enumitem}
\usepackage{subcaption}
\usepackage{xcolor}
\usepackage{graphicx}

\graphicspath{{images/}}

\title{NPE: Neural Personalized Embedding for Collaborative Filtering}

\author{
ThaiBinh Nguyen$^1$,
Atsuhiro Takasu$^{1,2}$
\\
$^1$ SOKENDAI (The Graduate University for Advanced Studies), Japan \\
$^2$ National Institute of Informatics, Japan\\
\{binh,takasu\}@nii.ac.jp
}

\begin{document}

\maketitle

\begin{abstract}
Matrix factorization is one of the most efficient approaches in recommender systems. However, such algorithms, which rely on the interactions between users and items, perform poorly for ``cold-users" (users with little history of such interactions) and at capturing the relationships between closely related items. To address these problems, we propose a neural personalized embedding (NPE) model, which improves the recommendation performance for cold-users and can learn effective representations of items. It models a user's click to an item in two terms: the \textit{personal preference} of the user for the item, and the \textit{relationships} between this item and other items clicked by the user. We show that NPE outperforms competing methods for top-N recommendations, specially for cold-user recommendations. We also performed a qualitative analysis that shows the effectiveness of the representations learned by the model.
\end{abstract}

\section{Introduction}
\label{sec:introduction}
In recent years, recommender systems have become a core component of online services. Given the ``historical activities" of a particular user (e.g., product purchases, movie watching, and Web page views), a recommender system suggests other items that may be of interest to that user. Current domains for recommender systems include movie recommendation (Netflix and Hulu), product recommendation (Amazon), and application recommendation (Google Play and Apple Store).

The historical activities of users are often expressed in terms of a user-item preference matrix whose entries are either \textit{explicit feedback} (e.g., ratings or like/dislike) or \textit{implicit feedback} (e.g., clicks or purchases). Typically, only a small part of the potential user-item matrix is available, with the remaining entries not having been recorded. Predicting user preferences can be interpreted as filling in the missing entries of the user-item matrix. In this setting, matrix factorization (MF) is one of the most efficient approaches to find the latent representations of users and items \cite{hu2008collaborative,salakhutdinov2008a,koren2008factorization}. To address the sparseness of the user-item matrix, additional data are integrated into MF as ``side information." This might include textual information for article recommendations \cite{conf_kdd_WangWY15,wang2011collaborative}, product images in e-commerce \cite{He:2016:VVB:3015812.3015834}, or music signals for song recommendations \cite{Oord:2013:DCM:2999792.2999907}. However, there are two major issues with these MF-based algorithms. First, these models are poor at modeling \textit{cold-users} (i.e., users who have only a short history of relevant activities). Second, because these models consider only user-item interactions, the item representations poorly capture the relationships among closely related items \cite{journals/cacm/Koren10}.

One approach to cold-user recommendation is to exploit user profiles. Such proposed models \cite{7929983,Li:2015:DCF:2806416.2806527} can learn user representations from their profiles (e.g., gender and age). In this way, these models can make recommendations to new users who have no historical activities, provided their user profiles are available. However, user profiles are often very noisy, and in many cases, they are simply not available. Another approach is item-similarity based models \cite{sarwar01itembased,Linden:2003:ARI:642462.642471}, which recommends items based on item--item similarity. The main issue of this approach is that it considers only the most recent click when making a recommendation, ignoring previous clicks. In addition, these models are not personalized.

In item representations learning, Item2Vec \cite{confrecsysBarkanK16} is an efficient model that borrows the idea behind word-embedding techniques \cite{confnipsMikolovSCCD13} for learning item representations. However, the main goal of Item2Vec is to learn item representations and it cannot be used directly for predicting missing entries in a user-item matrix. Furthermore, in making recommendations, Item2Vec is not personalized: it recommends items based on the similarities between items, computed using item representations, and ignores users' historical activities.

To address these problems, this paper proposes a neural personalized embedding (NPE) model that fuses item relationships for learning effective item representations in addition to improving recommendation quality for cold-users. NPE models a user's click on an item by assuming that there are two signals driving the click: the \textit{personal preference} of the user with respect to the item and the \textit{relationships} between this item and other items that the user has clicked.

To model the \textit{personal preference} term, we adopt the same approach as MF, which views the preference of a user for an item as the inner product of the corresponding factor vectors. To model the \textit{relationships} among items, we propose an \textit{item-embedding} model that generalizes the idea behind word-embedding techniques to click data. However, our item-embedding model differs from the word-embedding model in that the latter can only learn word representations. In contrast, our embedding model can both learn item representations and fill in the user-item matrix simultaneously.

\section{Related Work}
\label{sec:related_work}
\subsubsection{Matrix Factorization}
MF \cite{salakhutdinov2008a,koren2008factorization} is one of the most efficient ways to perform collaborative filtering. An MF-based algorithm associates each user with a latent feature vector of preferences and each item with a latent feature vector of attributes. Given prior ratings of users to items, MF learns the latent feature vectors of users and items and uses these vectors to predict missing ratings. To address sparseness in the user-item matrix, additional data about items/users are also used \cite{wang2011collaborative,Oord:2013:DCM:2999792.2999907,conf_kdd_WangWY15}.

Recently, the CoFactor \cite{confrecsysLiangACB16} and CEMF \cite{nguyen2017collaborative} models have been proposed. These models integrate item embedding into the MF model. They simultaneously decompose the preference matrix and the SPPMI matrix (the item--item matrix constructed from co-click information) in a shared latent space. However, in contrast to our proposed method, CoFactor and CEMF use co-click information to regularize the user-item matrix information, whereas NPE exploits co-click information for learning effective representations of items. In \cite{binh_iconip2017}, the author uses co-click information to address the data sparsity issue in rating prediction.

For cold-user recommendations, \cite{7929983} and \cite{Li:2015:DCF:2806416.2806527} proposed models that learn user presentations from user profiles. In \cite{7929983}, the user representations are learned from user profiles via a deep convolutional neural network for event recommendations, whereas \cite{Li:2015:DCF:2806416.2806527} has user representations being learned by an auto-encoder. Despite these models being very useful for new-user recommendations, the main issue remains that user profiles are not always available. Furthermore, many user profiles may be very noisy (e.g., users may not want to publish their real gender, age, or location), which leads to inaccurate representations of users.

\subsubsection{Embedding Models}
Word-embedding techniques \cite{confnipsMikolovSCCD13,conf/emnlp/LiZM15} have been applied successfully to many tasks in natural language processing. The goal of word embedding is to learn vector representations of words that capture the relationships with surrounding words. The assumption behind word embedding techniques is that words that occur in the same context are similar. To capture such similarities, words are embedded into a low-dimensional continuous space.

If an item is viewed as a word, and a list of items clicked by a user is a context window, we can map word embedding to recommender systems. \textit{Item2Vec} \cite{confrecsysBarkanK16} was introduced as a neural network-based item-embedding model. However, Item2Vec is not able to predict missing entries in a user-item matrix directly. Furthermore, in its recommendations, Item2Vec relies only on the last item, ignoring previous items that a user has clicked.

Exponential Family Embeddings (EFE) \cite{NIPS2016_6571}, a probabilistic embedding model that generalizes the spirit of word embedding to other kinds of data, which can be used for modeling clicks and learn item representations. However, EFE does not support for side information such as items' rich contents. In addition, EFE is not personalized.

\subsubsection{Item-based Models}
In item-based collaborative filtering \cite{sarwar01itembased,Ning:2012:SLM:2365952.2365983}, an item is recommended to a user based on the similarity between this item and the items that the user clicked in the past. In \cite{sarwar01itembased}, an item-- similarity matrix is constructed and is used directly to calculate the item similarities for recommendations. Previous work shows that the performance of this method is highly sensitive to the choice of similarity metric and data normalization \cite{Herlocker:2002:EAD:593967.594047}.

SLIM \cite{Ning:2011:SSL:2117684.2118303} is a recent model that identifies item similarity from the preference matrix by learning a sparse item-similarity matrix from the preference matrix. However, the disadvantage of SLIM is that it can only capture the relations between items that are co-clicked by at least one user. This will limit the capability of the model when applied to extremely sparse datasets. Furthermore, SLIM can only be used to predict the missing entries of the user-item matrix and cannot be used for learning effective representations of items.

\section{NPE: Neural Personalized Embedding}
We propose NPE, a factor model that explains users' clicks by capturing the preferences of users for items and the relationships between closely related items. We will describe the model and how to learn the model parameters.

\label{sec:proposed_model}
\subsection{Problem Formulation}
Each entry $r_{u,i}$ in the user-item preference matrix $\mathbf{R}$ has one of two values $0$ or $1$, such that $r_{u,i}=1$ if user $u$ has clicked item $i$ and $r_{u,i}=0$ otherwise. We assume that $r_{u,i}=1$ indicates that user $u$ prefers $i$, whereas $r_{u,i}=0$ indicates that this entry is non-observed (i.e., a missing entry).

Given a user $u$ and the set of items that $u$ previously interacted, our goal is to predict a list of items that $u$ may find interesting (top-N recommendations).

The notations used in this paper are defined in Table \ref{tab:notations}.
\begin{table}[tb]
  \centering
  \renewcommand{\arraystretch}{1.2}
  \begin{tabularx}{\linewidth}{c|X}
    Notation&Meaning\\
    \hline
    $N, M$ & the number of users and items, respectively\\
    $\mathbf{R}$ & the user-item matrix (e.g., click matrix)\\
    $\mathbf{r}_u$ & the observation data for user $u$ (i.e., the row corresponding to user $u$ of matrix $\mathbf{R})$\\
    $D$ & the dimensionality of the embedding space\\
    $H$ & the dimensionality of the user input vector\\
    $L$ & the dimensionality of the item input vector\\
    $\mathbf{x}_u$ & the input vector of user $u$, $x_u\in\mathbb{R}^H$\\
    $\mathbf{y}_i$ & the input vector of item $i$, $y_i\in\mathbb{R}^L$\\
    $\mathbf{H}$ & the user embedding matrix, $\mathbf{H}\in\mathbb{R}^{H\times D}$\\
    $\mathbf{W}$ & the item-embedding matrix, $\mathbf{W}\in\mathbb{R}^{L\times D}$\\
    $\mathbf{V}$ & the item context matrix, $\mathbf{V}\in\mathbb{R}^{L\times D}$\\
    $\mathbf{h}_u$ & the embedding vector of user $u$, $\mathbf{h}_u\in\mathbb{R}^D$\\
    $\mathbf{w}_i$ & the embedding vector of item $i$, $\mathbf{w}_i\in\mathbb{R}^D$\\
    $\mathbf{v}_i$ & the context vector of item $i$, $\mathbf{v}_i\in\mathbb{R}^D$\\
    $\bm{\Theta}$ & The set of all model parameters\\
    $\Omega(.)$ & The regularization term\\
    $\mathbf{c}_{u,i}$ & the set of items that user $u$ clicked, excluding $i$ (the \textit{context} items)\\
    $\mathcal{D}^+$ & the set of positive examples, $\mathcal{D}^+=\{(u,i)|r_{u,i}=1\}$\\
    $\mathcal{D}^-$ & the set of negative examples, which is obtained by sampling from zero entries of matrix $\mathbf{R}$\\
\end{tabularx}
  \caption{The notations used throughout the paper.}
  \label{tab:notations}
\end{table}

\subsection{Model Formulation}
We denote the observations for user $u$ as:
\begin{equation}
\mathbf{r}_u=(r_{u,1}, r_{u,2}, \dots, r_{u,M}).
\end{equation}

NPE models the probability of each observation conditioned on user $u$ and its context items as:
\begin{equation}
\label{eq:likelihood_one}
  p(r_{u,i}=1|u, \mathbf{c}_{u,i}),
\end{equation}

This equation captures the intuition behind the model, namely that the conditional distribution of whether user $u$ clicks on item $i$ is governed by two factors: (1) the personal preference of user $u$ for item $i$, and (2) the set of items that $u$ has clicked (i.e., $\mathbf{c}_{u,i}$).

The likelihood function for the entire matrix $\mathbf{R}$ is then formulated as:
\begin{equation}
  \label{eq:likelihood_entire}
  p(\mathbf{R})=\prod_{u=1}^N\prod_{i=1}^{M}p(r_{u,i}|u,\mathbf{c}_{u,i}).
\end{equation}

The conditional probability expressed in Eq. \ref{eq:likelihood_one} is implemented by a neural network. This neural network connects the input vectors of user $u$, item $i$, and context items $\mathbf{c}_{u,i}$ to their hidden representations as:
\begin{align}
  \mathbf{h}_u&=\mathbf{f}(\mathbf{x}_u^\top\mathbf{H}),\\
  \mathbf{w}_i&=\mathbf{f}(\mathbf{y}_i^\top\mathbf{W}),\\
  \mathbf{v}_{\mathbf{c}_{u,i}}&=\mathbf{f}(\sum_{j\in\mathbf{c}_{u,i}}\mathbf{y}_i^\top\mathbf{V}),
\end{align}
where $\mathbf{f}(.)$ is an activation function such as ReLU.

Note that there are two hidden representations associated with item $i$: the \textit{embedding vector} $\mathbf{w}_i$ and the \textit{context vector} $\mathbf{v}_i$, which have different roles. Whereas $\mathbf{w}_i$ accounts for the attributes of item $i$, $\mathbf{v}_i$ accounts for specifying the items that appear in its context.

We can then define the conditional probability in Eq. \ref{eq:likelihood_one} via the hidden representations as:
\begin{equation}
\label{eq:conditional_probability_function}
p(r_{u,i}=1|u, \mathbf{c}_{u,i})=\sigma(\mathbf{h}_u^\top\mathbf{w}_i+\mathbf{w}_i^\top\mathbf{v}_{\mathbf{c}_{u,i}}).
\end{equation}
Note that the $\sigma(.)$ function on the right side of Eq. \ref{eq:conditional_probability_function} comprises two terms: the first term $\mathbf{h}_u^\top\mathbf{w}_i$ accounts for how user $u$ prefers item $i$, whereas the second term $\mathbf{w}_i^\top\mathbf{v}_{\mathbf{c}_{u,i}}$ accounts for the compatibility between item $i$ and the items that $u$ has already clicked.

From Eq. \ref{eq:conditional_probability_function}, we can also obtain the probability that $r_{u,i}=0$ as:
\begin{equation}
\label{eq:conditional_distribution_negative}
\begin{aligned}
p(r_{u,i}=0|u, \mathbf{c}_{u,i})&=1-\sigma(\mathbf{h}_u^\top\mathbf{w}_i+\mathbf{w}_i^\top\mathbf{v}_{\mathbf{c}_{u,i}})
\end{aligned}
\end{equation}

The conditional probability functions in Eqs. \ref{eq:conditional_probability_function} and \ref{eq:conditional_distribution_negative} can be summarized in a single conditional probability function as:
\begin{equation}
    p(r_{u,i}=r|u, \mathbf{c}_{u,i})=
    \begin{cases}
      \hat{\mu}_{u,i}, & \text{if}\ r=1, \\
      1-\hat{\mu}_{u,i}, & \text{if}\ r=0,
    \end{cases}
\end{equation}
where $\hat{\mu}_{u,i}=\sigma(\mathbf{h}_u^\top\mathbf{w}_i+\mathbf{w}_i^\top\mathbf{v}_{\mathbf{c}_{u,i}})$.

\subsection{The Model Architecture}
\label{sub:model_architecture}
\begin{figure}[b]
\centering
  \includegraphics[width=0.90\linewidth]{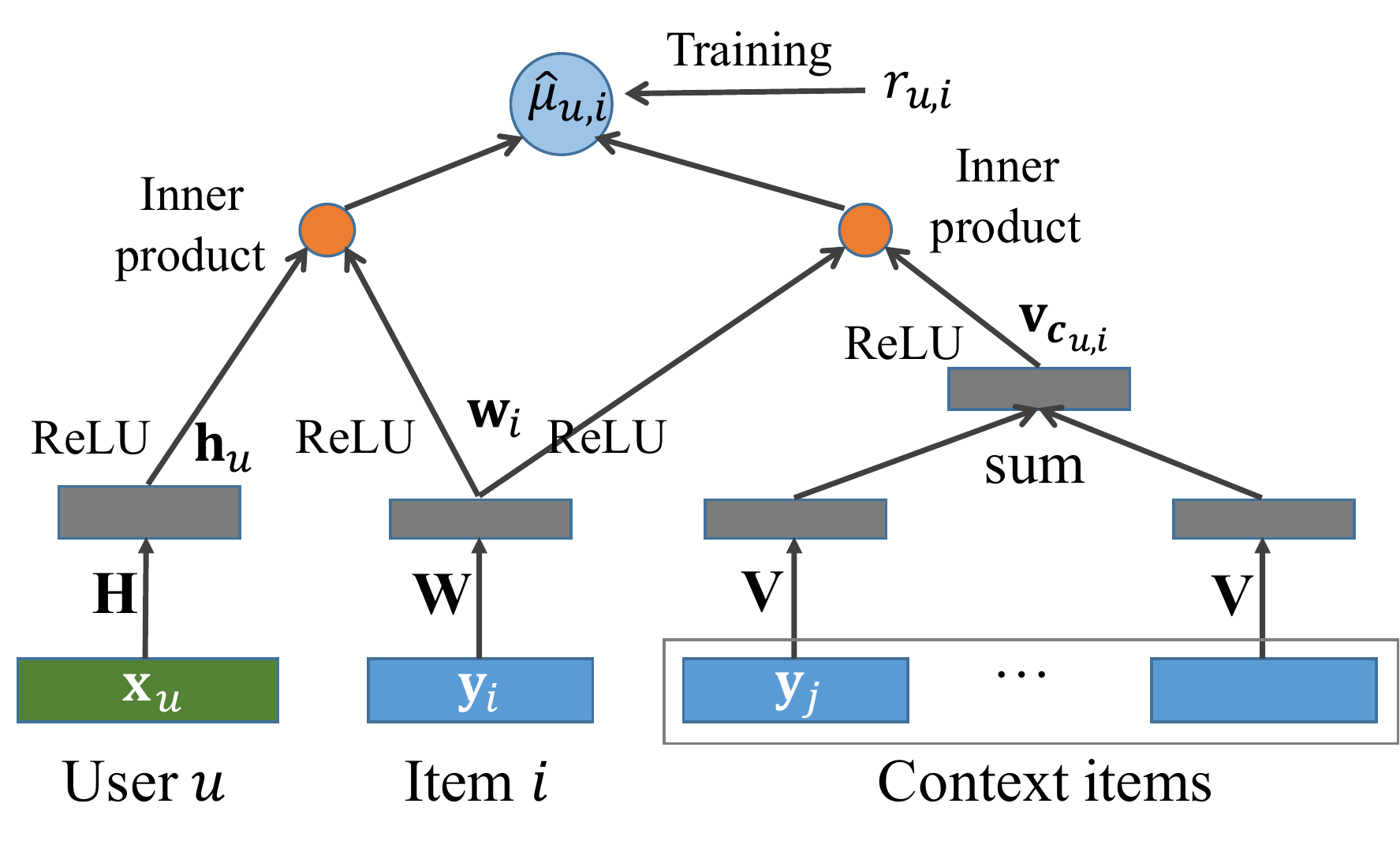}
  \caption{The architecture of NPE.}
  \label{fig:the_model}
\end{figure}

The architecture of NPE is shown in Fig. \ref{fig:the_model} as a multi-layer neural network. The first layer is the input layer which specifies the input vectors of (1) a user $u$, (2) a candidate item $i$, and (3) the context items. Above this is the second layer (the embedding layer), which connects to the input layer via connection matrices $\mathbf{H}$, $\mathbf{W}$, and $\mathbf{V}$.
Above the embedding layer, two terms are calculated: the \textit{personal preference} of user $u$ for item $i$ and the \textit{relationship} between $i$ and the context items. Finally, the model combines these two terms to compute the output, which is the probability that $u$ will click $i$.

Note that, the input layer accepts a wide range of vectors that describe users and items such as one-hot vector or content feature vectors obtained from side information. With such a generic input vectors, our method can address the cold-start problem by using content feature vectors as input vectors for users and items. Since this work focuses on the pure collaborative filtering setting, we use only the identities of users and items in the form of one-hot vectors as input vectors. Investigating the effectiveness of using content feature vectors, is left for future work.

\subsection{Objective Function}
Given an observed matrix $\mathbf{R}$, our goal is to learn the model parameters $\bm{\Theta}$ that maximize the likelihood function in Eq. \ref{eq:likelihood_entire}. However, instead of modeling all zero entries, we only model a small subset of such entries by picking them randomly (negative sampling). This gives:
\begin{align}
\label{eq:likelihood_negative_sampled}
\begin{aligned}
p(\mathbf{R}) = \prod_{(u,i)\in \mathcal{D}^+} p(r_{u,i}|u, \mathbf{c}_{u,i})\prod_{(u,i)\in \mathcal{D}^-} p(r_{u,i}|u, \mathbf{c}_{u,i}).
\end{aligned}
\end{align}

Maximizing the likelihood in Eq. \ref{eq:likelihood_negative_sampled} is equivalent to minimizing the following loss function (its negative log function):
\begin{align}
\begin{aligned}
\label{eq:loss_function}
\mathcal{L}(\bm{\Theta})&=-\sum_{(u,i)\in \mathcal{D}^+}\log\hat{\mu}_{ui}-\sum_{(u,i)\in \mathcal{D}^-}\log(1-\hat{\mu}_{ui})\\
&\quad + \lambda\Omega(\bm{\Theta}),
\end{aligned}
\end{align}
where $\hat{\mu}_{u,i}=\sigma(\mathbf{h}_u^\top\mathbf{w}_i+\mathbf{w}_i^\top\mathbf{v}_{\mathbf{c}_{u,i}})$.

This loss function is known as the \textit{binary cross-entropy}.

\subsection{Model Training}
We adopt the Adam technique (a mini-batch stochastic gradient descent approach) \cite{DBLP:journals/corr/KingmaB14}. We do not perform negative sampling in advance, which can only produce a fixed set of negative samples. Instead, we perform negative sampling with each epoch, which enables diverse sets of negative examples to be used. The algorithm is summarized in Algorithm \ref{alg:npe}.

\begin{algorithm}[t]
\SetAlgoLined
\SetKwInOut{Input}{Input}
\SetKwInOut{Output}{Output}
\Input{\begin{itemize}[label={--}]
        \item $\mathbf{R}$: User-item preference matrix
        \item $n$: number of negative samples per positive example
        \end{itemize}}
\Output{$\bm{\Theta}=\{\mathbf{H}, \mathbf{W},\mathbf{V}\}$}
 Initialization: sample $\mathbf{H}, \mathbf{W},\mathbf{V}$ from Gaussian distributions\\
 \For{epoch=1 \dots T}{
  Sample negative examples $\mathcal{D}^-$\\
  $\mathcal{D}=\mathcal{D}^+\cup \mathcal{D}^-$\\
  $\mathcal{O} = \text{Shuffle}(\mathcal{D})$\\
  \For{t=1 \dots \text{\# of mini-batches}}{
    $\mathcal{B}=\text{next-mini-batch}(\mathcal{O})$\\
    \text{Backprop}($\bm{\Theta}, \mathcal{B}$)
  }
  }
 \caption{NPE($\mathbf{R}, n$). Backprop is the back-propagation procedure for updating network weights.}
\label{alg:npe}
\end{algorithm}

\subsection{Connections with Previous Models}
\subsubsection{NPE vs. MF}
In the conditional probability in Eq. \ref{eq:conditional_probability_function}, we can see that the $\sigma(.)$ function is a combination of two terms: (1) user preference and (2) item relationship. If the second term is removed, NPE will reduce to an original MF method.
\subsubsection{NPE vs. Word Embedding}
Similarly, if we remove the first element of $\sigma(.)$ in Eq. \ref{eq:conditional_probability_function}, NPE will model only the relationship among items. If we view each item as a word, and the set of items that a user clicked as a sentence, the model becomes similar to a word-embedding model. However, our embedding model differs in that word-embedding techniques can only learn word (item) representations and cannot fill the user-item matrix directly. In contrast, our embedding model can learn effective item representations while predicting the missing entries in the user-item matrix.

\section{Empirical Study}
\label{sec:empirical_study}
We have studied the effectiveness of NPE both quantitatively and qualitatively. In our quantitative analysis, we compared NPE with state-of-the-art methods on top-N recommendation task, using real-world datasets. We also performed a qualitative analysis to show the effectiveness of the item representations.
\subsection{Datasets}
We used three real-world datasets whose sizes varied from small to large-scale, from different domains. First, \textit{Movielens 10M} (ML-10m) is a dataset of user-movie ratings, collected from MovieLens, an online film service. Next, \textit{Online Retail} \cite{Chen2012_OnlineRetail} is a dataset of online retail transactions that contains all transactions from Dec 1, 2010 to Dec 9, 2011 for an online retailer. Finally, \textit{TasteProfile} is a dataset of counts of song plays by users, as collected by Echo Nest.\footnote{http://the.echonest.com/}

\begin{table}[tb]
  \begin{tabular}{ccccc}
    \toprule
    &ML-10m&OnlineRetail&TasteProfile\\
    \midrule
    \hline
    \#users &  58,059& 3,705& 211,830\\
    \#items &  8,484& 3,644& 22,781\\
    \#clicks  &3,502,733 & 235,472 & 10,054,204\\
    \% clicks  & 0.71\% & 1.74\% &0.21\%\\
  \bottomrule
\end{tabular}
  \caption{Statistical information about the datasets.}
  \label{tab:data_information}
\end{table}

\subsection{Experiment Setup}
\subsubsection{Data Preparation}
For the ML-10m, we binarized the ratings, thresholding at 4 or above; for TasteProfile and OnlineRetail, we binarized the data and interpreted them as implicit feedback.
Statistical information about the datasets is given in Table \ref{tab:data_information}.

We partitioned the data into three subsets, using 70\% of the data as the training set, 10\% as the validation set, and the remaining 20\% as the test set (ground truth).

\subsubsection{Evaluation Metrics} After training the models on the training set, we evaluated the accuracy of their top-N recommendations using the test set. We used the rank-based metrics Recall@$n$ and nDCG@$n$, which are common metrics in information retrieval, for evaluating the accuracy of the top-N recommendations. (We did not use ``Precision" because it is difficult to evaluate, given that a zero entry can imply either that the user does not like the item or does not know about the item).

\subsubsection{Competing Methods} We compared NPE with the following competing methods:
\begin{itemize}
  \item Bayesian personalized ranking (\textbf{BPR}) \cite{bpr2009}: an algorithm that optimizes the MF model with a pair-wise ranking loss
  \item Neural collaborative filtering (\textbf{NeuCF}) \cite{He:2017:NCF:3038912.3052569}: a generalization of an MF method in which the inner product of user and item feature vectors are replaced by a deep neural network
  \item Sparse linear model (\textbf{SLIM}) \cite{Ning:2011:SSL:2117684.2118303}: a state-of-the-art method for top-N recommendations, which is based on the similarities between items.
\end{itemize}

\subsection{Implementation Details}
Since neural networks are prone to overfitting, we apply a dropout after the hidden representation layer. The dropout rate is tuned for each dataset. We use early stopping to terminate the training process if the loss function does not decrease on the validation set for five epochs. The weights for the matrices $\mathbf{H}$, $\mathbf{W}$, and $\mathbf{V}$ are initialized as normal distributions. The size of each mini-batch was 10,000.


\subsection{Experimental Results}
\subsubsection{Top-N Recommendations} Table \ref{tab:top_n_comparison} summarizes the Recall@20 and nDCG@20 for each model. Note that NPE significantly outperforms the other competing methods across all datasets for both Recall and nDCG. We emphasize that all methods used the same data. However, NPE benefits from capturing the compatibility between each item and other items picked by the same users.
\begin{table*}[ht]
\centering
\begin{tabular}{l|cc||cc||cc}
    \hline
    \multirow{2}{5em}{Methods} &
      \multicolumn{2}{c||}{ML-10m}
      &
      \multicolumn{2}{c||}{OnlineRetail}
      &
      \multicolumn{2}{c}{TasteProfile}\\
    & Re@20 & nDCG@20 & Re@20 & nDCG@20 & Re@20 & nDCG@20\\
    \hline
    \hline
    SLIM & 0.1342 & 0.1289 & 0.2085 & 0.1015 & 0.1513 & 0.1422\\
    BPR & 0.1314 & 0.1253 & 0.2137 & 0.0943 & 0.1598 & 0.1398\\
    NeuCF & 0.1388 & 0.1337 & 0.2199 & 0.0911 & 0.1609 & 0.1471\\
    NPE (our) & \textbf{0.1497} & \textbf{0.1449} & \textbf{0.2296} & \textbf{0.1742} & \textbf{0.1788} & \textbf{0.1594}\\
    \hline
  \end{tabular}
\caption{Recall and nDCG for three datasets, with embedding size $D=64$ and negative sampling ratio $n=4$.}
\label{tab:top_n_comparison}
\end{table*}

In Table \ref{tab:impact_top_n}, we summarize Recall@20 values for the four methods when different numbers of items were to be recommended. From these results, we can see that NPE consistently outperformed the other methods at all settings. The differences between NPE and the other methods are more pronounced for small numbers of recommended items. This is a desirable feature because we often only a consider a small number of top items (e.g., top-$5$ or top-$10$).
\begin{table*}[ht]
\centering
\begin{tabular}{l|ccc||ccc||ccc}
    \hline
    \multirow{2}{5em}{Methods} &
      \multicolumn{3}{c||}{ML-10m}
      &
      \multicolumn{3}{c||}{OnlineRetail}
      &
      \multicolumn{3}{c}{TasteProfile}\\
    & Re@5 & Re@10 & Re@20  & Re@5 & Re@10 & Re@20 & Re@5 & Re@10 & Re@20\\
    \hline
    \hline
    SLIM & 0.1284& 0.1298& 0.1342 & 0.0952& 0.1311 & 0.2085 & 0.1295& 0.1304& 0.1513\\
    BPR & 0.1254 &	0.1261	& 0.1314 & 0.0859 & 0.1222	& 0.2137 & 0.1307 & 0.1311 & 0.1598\\
    NeuCF & 0.1347&	0.1363 & 0.1388 & 0.0871 &	0.1274	& 0.2199 & 0.1342 & 0.1356 & 0.1609\\
    NPE (our) & \textbf{0.1451} & \textbf{0.1487} & \textbf{0.1497} & \textbf{0.1392} & \textbf{0.1667} & \textbf{0.2296} & \textbf{0.1428} & \textbf{0.1523} & \textbf{0.1788}\\
    \hline
  \end{tabular}
\caption{Recall for different numbers of items to be recommended, with embedding size $D=64$ and negative sampling ratio $n=4$.}
\label{tab:impact_top_n}
\end{table*}

\subsubsection{The Performance on Cold-Users}
We studied the performance of the models for users who had few historical activities. To this end, we partitioned the test cases into three groups, according to the number of clicks that each user had. The \textit{Low} group's users had less than $10$ clicks, the \textit{Medium} group's users had $10 \sim 20$ clicks, and the \textit{High} group's users had more than $20$ clicks.

\begin{figure}[tb]
\centering
\begin{subfigure}{0.45\linewidth}
\captionsetup{skip=0pt}
\includegraphics[width=1.0\linewidth]{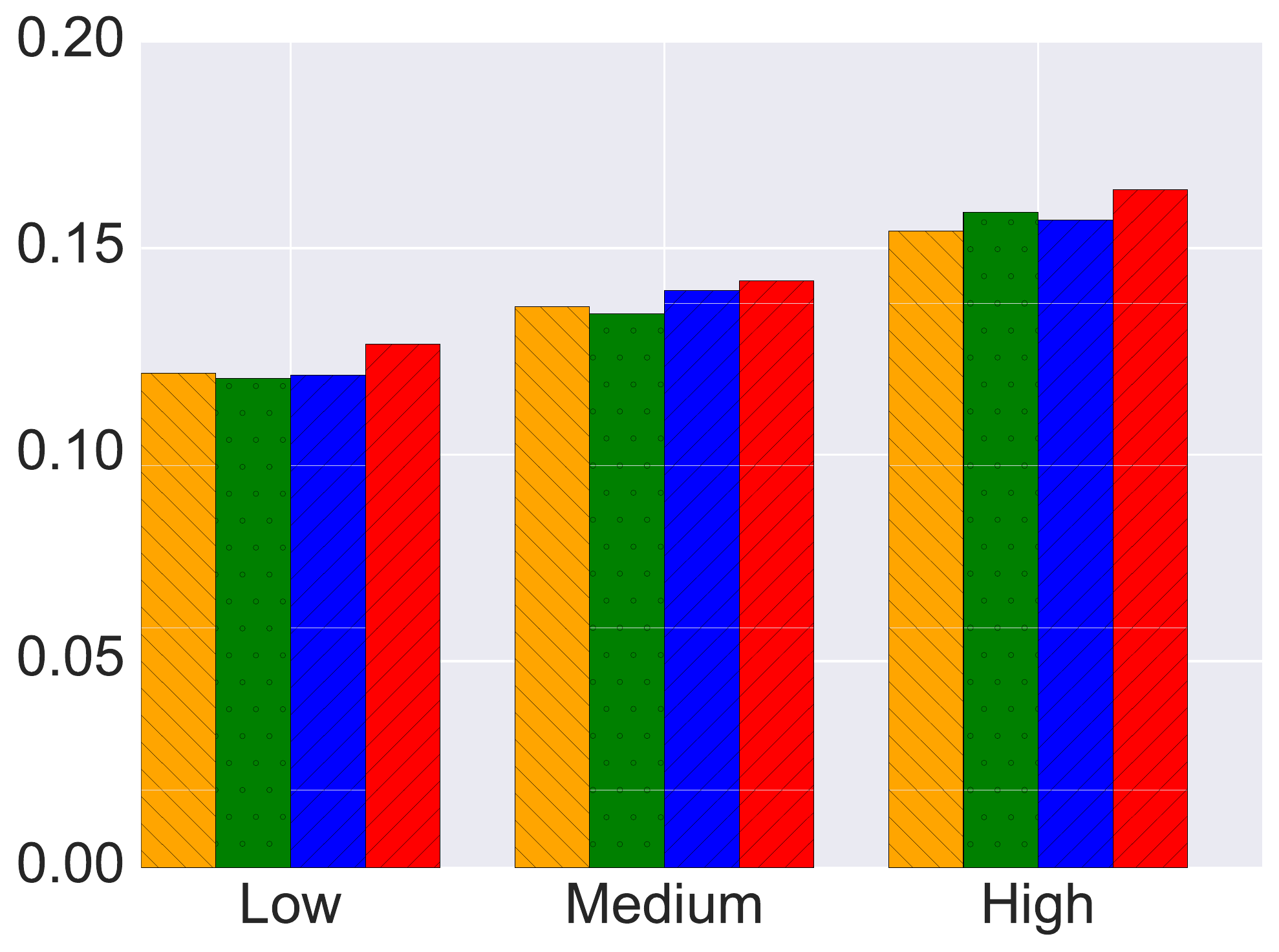}
\caption{ML-10m}\label{fig:ml10m_recall_dim}
\end{subfigure}
\begin{subfigure}{0.45\linewidth}
\captionsetup{skip=0pt}
\includegraphics[width=1.0\linewidth]{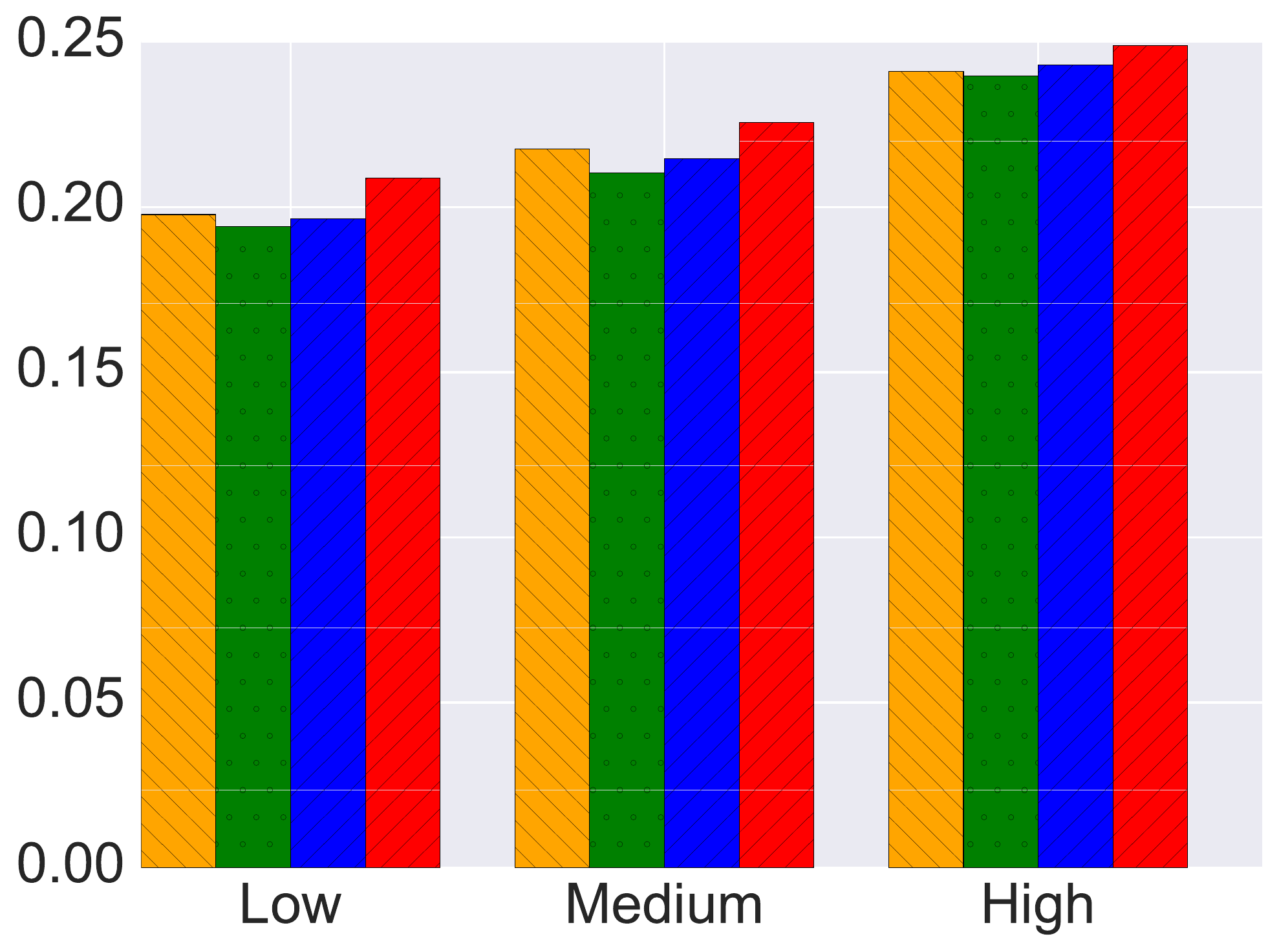}
\caption{OnlineRetail}\label{fig:onlineretail_recall_dim}
\end{subfigure}
\begin{subfigure}{1\linewidth}
  \centering
  \includegraphics[width=0.8\linewidth]{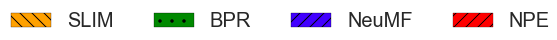}
\end{subfigure}
\caption{Recall@$20$ for different groups of users.}
\label{fig:cold_start_users_recall}
\end{figure}

Fig. \ref{fig:cold_start_users_recall} shows the breakdown of Recall@20 in terms of user activity in the training set for the ML-10m and OnlineRetail. Although the details varied across datasets, the NPE model outperformed the other methods for all three groups of users. The differences between NPE and the other methods are much more pronounced for users who have fewest clicks. This is to be expected because, for such users, NPE captures the item relations when making recommendations.

\subsubsection{Effectiveness of the Item Representations} We evaluated the effectiveness of item representations by investigating how well the representations capture the \textit{item similarity} and items that are often \textit{purchased together}.

\begin{figure}[tb]
\centering
  \includegraphics[width=1.0\linewidth]{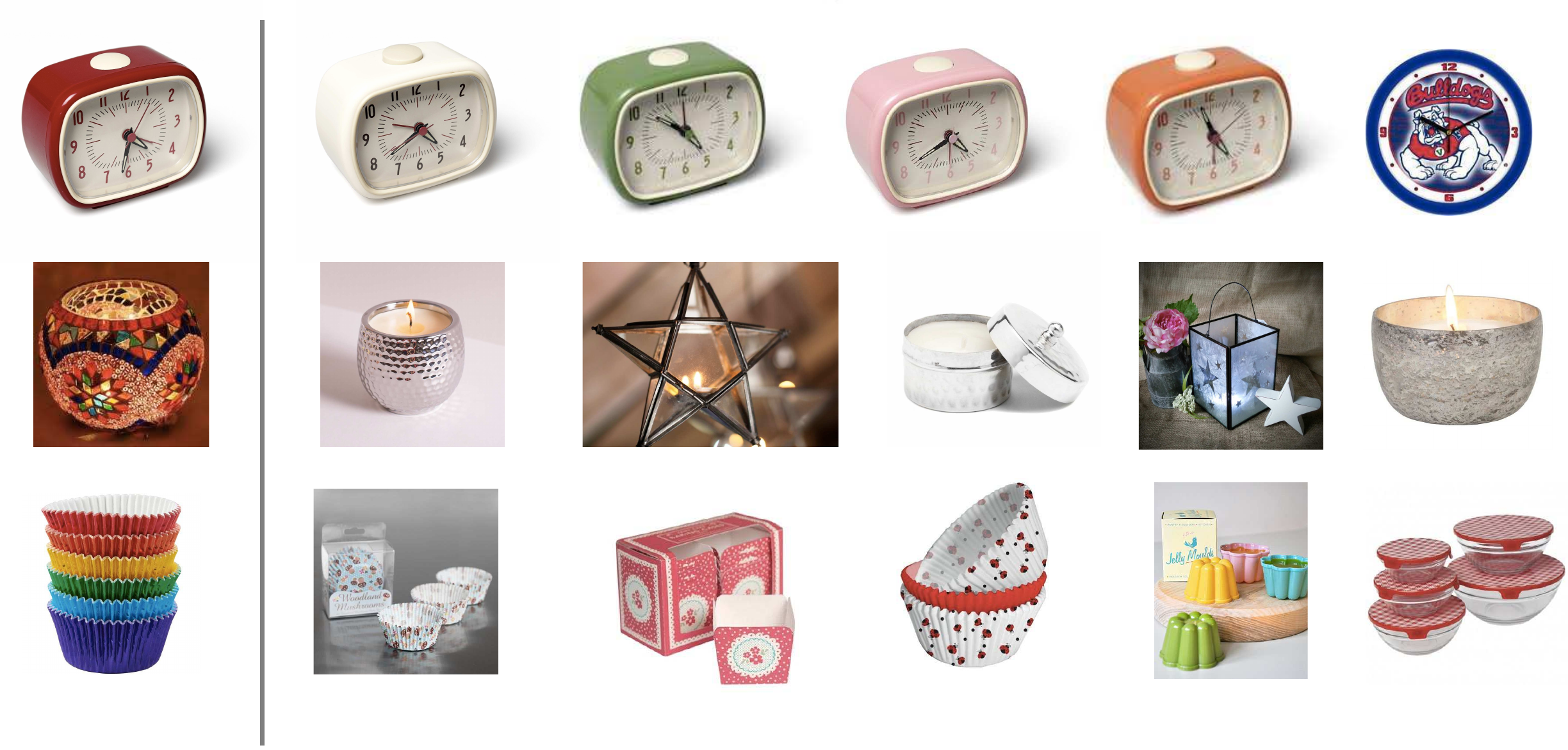}
  \caption{Top-5 similar items for a given item. In each row, the given item is at the left and the top-5 similar items are to its right.}
\label{fig:similar_item}
\end{figure}

\textbf{Similar items:} The similarity between two items is defined as the cosine distance between their embedding vectors. Fig. \ref{fig:similar_item} shows three examples of the top-5 most similar items to a given item in the OnlineRetail dataset. We can see that the items' embedding vectors effectively capture the similarity of the items. For example, in the first row, given a \textit{red alarm clock}, four of its top-5 similar items are also alarm clocks.

\begin{figure}[!tb]
\centering
  \includegraphics[width=1.0\linewidth]{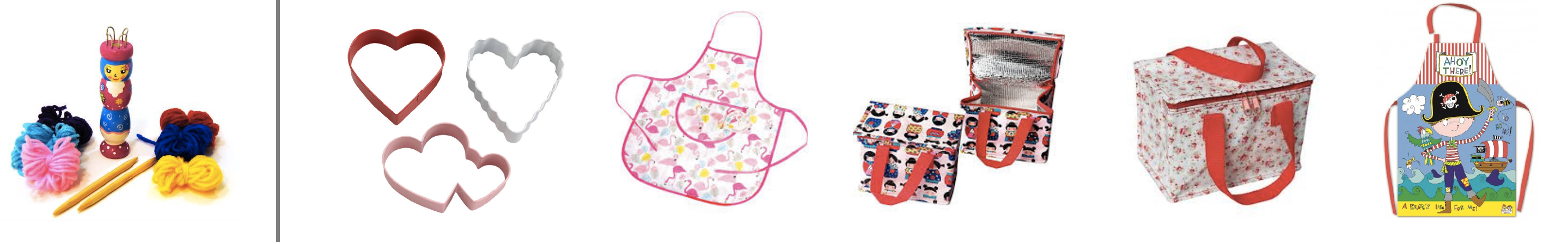}
  \caption{Top-5 items that are likely to be bought together with a given item. The given item is at the left and its top-5 most similar items are to its right.}
\label{fig:co_purchased_item}
\end{figure}

\textbf{Items that are often purchased together:} NPE can also identify items that are often purchased together. To assess if two items are often purchased together, we calculate the inner product of one item's embedding vector $\mathbf{w}_i$ and the other's context vector $\mathbf{v}_j$. A high value of this inner product indicates that these two items are often purchased together. Fig. \ref{fig:co_purchased_item} shows an example of items that tend to be purchased together with the given item. Here, we see that buying a \textit{knitting Nancy}, a child's toy, might accompany the purchase of other goods for children or for a household.

\subsubsection{Sensitivity Analysis}
We also studied the effect of the hyper-parameters on the models' performance.

\textbf{Impact of the embedding size:} To evaluate the effects of the dimensionality of the embedding space on the top-N recommendations, we varied the embedding dimension $D$ while fixing the other parameters. Table \ref{tab:impact_embedding_size} summarizes the Recall@$20$ for NPE on the three datasets for various embedding sizes: $D=\{8, 16, 32, 64, 128, 256\}$. We can see that the larger embedding sizes seem to improve the performance of the models. The optimal embedding size for OnlineRetail is $D=64$ and, for ML-10m and TasteProfile is $n=128$.

\begin{table}[tb]
\centering
\begin{tabular}{l|c||c||c}
    \hline
    \multirow{2}{2em}{$D$} &
      \multicolumn{1}{c||}{ML-10m}
      &
      \multicolumn{1}{c||}{OnlineRetail}
      &
      \multicolumn{1}{c}{TasteProfile}\\
    & Re@20 & Re@20 & Re@20\\
    \hline
    \hline
    8 & 0.1428& 0.1187& 0.0987\\
    16 & 0.1451& 0.1596& 0.1142\\
    32 & 0.1441& 0.1950& 0.1509\\
    64 & \textbf{0.1497}& \textbf{0.2296}& 0.1788\\
    128 & 0.1482& 0.2284& \textbf{0.1992}\\
    256 & 0.1459& 0.2248& 0.1985\\
    \hline
  \end{tabular}
	\caption{Recall@20 for various embedding sizes, with negative sampling ratio $n=4$.}
	\label{tab:impact_embedding_size}
\end{table}

\textbf{Impact of the negative sampling ratio:} During the training of NPE, we sampled negative examples. We studied the effect of the negative sampling ratio $n$ on the performance of NPE by fixing the embedding size $D=32$ and evaluating Recall@20 for $n=\{1,2,4,5,8,12,16,20\}$. From Table \ref{tab:impact_negative_ratio}, we note that when $n$ increases, the performance also increases up to a certain value of $n$. The optimal negative sampling ratios are $n=\{4,5\}$ for OnlineRetail and $n=8$ for ML-10m and TasteProfile. This is reasonable because ML-10m and TasteProfile, being larger than OnlineRetail, will need more negative examples.
\begin{table}[tb]
\centering
\begin{tabular}{l|c||c||c}
    \hline
    \multirow{2}{2em}{$n$} &
      \multicolumn{1}{c||}{ML-10m}
      &
      \multicolumn{1}{c||}{OnlineRetail}
      &
      \multicolumn{1}{c}{TasteProfile}\\
    & Re@20 & Re@20 & Re@20\\
    \hline
    \hline
    1 & 0.1392& 0.1608& 0.1243\\
    2 & 0.1418& 0.1795& 0.1451\\
    4 & 0.1441& 0.1950& 0.1509\\
    5 & 0.1478& \textbf{0.1952}& 0.1585\\
    8 & \textbf{0.1563}& 0.1941& \textbf{0.1621}\\
    12 & 0.1531& 0.1937& 0.1615\\
    16 & 0.1524& 0.1925& 0.1603\\
    20 & 0.1496& 0.1908& 0.1598\\
    \hline
  \end{tabular}
	\caption{Recall@20 for different negative sampling ratios, with a fixed embedding size $D=32$.}
	\label{tab:impact_negative_ratio}
\end{table}

\section{Conclusions and Future Work}
\label{sec:discussion}
We propose NPE, a neural personalized embedding model for collaborative filtering, is effective in making recommendations to cold-users and for learning item representations. Our experiments have shown that NPE can outperform competing methods with respect to top-N recommendations in general, and to cold-users in particular. Our qualitative analysis also demonstrated that item representations can capture effectively the different kinds of relationships between items.

One future direction will be to study the effectiveness of the model when using available side information about items.
We also aim to investigate different negative sampling methods for dealing with zero values in the user-item matrix.

\bibliographystyle{named}
\bibliography{ijcai18}

\end{document}